\documentstyle[]{mn} \voffset -0.5in

\begin{document}
\title{WR 146 - Observing the OB-type companion}
\author[S.M.~Dougherty et al.]
{S.M.~Dougherty$^{1,2}$, P.M. Williams$^{3}$ and D.L. Pollacco$^{4}$\\
$^1$ Dominion Radio Astrophysical Observatory, P.O. Box 248,
White Lake Rd, Penticton, British Columbia V2A 6K3, Canada\\
$^2$ Dept. of Physics and Astronomy, University of Calgary,
2500 University Dr. NW., Calgary, Alberta T2N 1N4, Canada\\
$^3$ Institute for Astronomy, University of Edinburgh,
Blackford Hill, Edinburgh EH9 3HJ\\
$^4$ Department of Pure and Applied Physics, Queen's University,
Belfast, BT7 1NN\\
}

\date{Originally submitted 1999 August 26; resubmitted 1999 December 2
and 2000 February 4}
\maketitle

\begin{abstract}
We present new radio and optical observations of the colliding-wind
system WR\,146 aimed at understanding the nature of the companion to
the Wolf-Rayet star and the collision of their winds.  The radio
observations reveal emission from three components: the WR stellar
wind, the non-thermal wind-wind interaction region and, for the first
time, the stellar wind of the OB companion.  This provides the unique
possibility of determining the mass-loss rate and terminal wind
velocity ratios of the two winds, independent of distance.
Respectively, these ratios are determined to be $0.20\pm0.06$ and
$0.56\pm0.17$ for the OB-companion star relative to the WR star.  A
new optical spectrum indicates that the system is more luminous than
had been believed previously. We deduce that the ``companion'' cannot
be a single, low luminosity O8 star as previously suggested, but is
either a high luminosity O8 star, or possibly an O8+WC binary system.
\end{abstract}


\section{Introduction}

Wolf-Rayet stars are surrounded by dense stellar winds giving rise to
free-free emission extending from IR to radio wavelengths. Typically,
this emission is characterised by a power-law spectrum of the form
$S_\nu\propto \nu^{\alpha}$, with values of the spectral index $\alpha
\sim +0.7-+0.8$, and radio brightness temperatures $\sim10^4$~K. A small
number of WR stars have radio emission that exhibits quite different
properties: negative spectral indices and brightness temperatures
$\sim10^6$~K or higher, properties that are characteristic of
non-thermal emission. WR\thinspace146 is a member of this group, which
includes WR\thinspace125, WR\thinspace140 and WR\thinspace147.

The radio emission from WR\thinspace146 was first resolved in high
resolution observations with MERLIN (Dougherty et al. 1996, hereafter
Paper I). These 5-GHz observations revealed two components, N$_5$ and
S$_5$, separated by $\sim120$ milli-arcseconds (mas). The flux of
S$_5$ was consistent with that estimated from extrapolation of the
IR-millimetre spectrum arising from the free-free emitting envelope
around the WR star. The brightness temperature of N$_5$ ($\sim10^6$K)
identified the nature of the emission from this component as
non-thermal. An optical spectrum showed evidence for absorption lines
at H$\delta$ and H$\gamma$, which we attributed to an early-type
companion to the WR star.  This led us to hypothesize in Paper I that
the non-thermal emission arose from a population of relativistic
electrons, accelerated in a wind-wind collision region where the wind
of the WR star and the companion interacted (e.g. Eichler \& Usov
1993). To be consistent with such a model, we suggested the companion
lay at the same position angle as N$_5$ from S$_5$, but slightly
further away from the WR star.

The presence of a companion was confirmed in optical imaging with the
Hubble Space Telescope (HST) by Niemela et al. (1998). They observed
two stars, WR146A and B (hereafter S$_{\rm O}$ and N$_{\rm O}$
respectively), at the same position angle as the radio sources but
separated by $\sim 168$ mas. Under the assumption that the southern
sources in both the HST and MERLIN images are coincident, these
observations place the non-thermal source between the two stellar
images, strongly supporting wind-wind collision as the origin of the
non-thermal emission. Its position relative to the two stellar
components ($\sim 120$ mas from S$_O$ and $\sim$ 48 mas from N$_O$) is
where the dynamical pressure of the two stellar winds is
balanced. This indicates that the momentum of S$_O$'s stellar wind is
$\sim$ 0.1 times that of N$_O$. With the wind velocity of the WC star
in WR\,146 ($\sim$ 2900 km s$^{-1}$, Eenens \& Williams 1994) being
greater than that of a typical OB star, and the expectation that the
mass-loss rate of a WR star would be greater than that of an OB star,
this strongly supports the identification of S$_O$ with the WC6
star in Paper I and identifies N$_O$ with an OB companion having a
lesser wind momentum.  The momentum ratio lead Niemela et al. to infer
that the wind momentum of the companion was more appropriate to a star
of early O or Of type than a late O-type main-sequence star. Taking
account of the photometry, they suggested an O6-O5 V-III spectral type
for the companion.

The photometry by Niemela et al. showed the two stars to be equally
bright in B but that S$_{\rm O}$ was redder than N$_{\rm O}$ in
(B--V) and (U--B). The (near-zero) magnitude difference between the
two blue images falls between two very different estimates of the
WR:O light ratio deduced from spectra of WR\,146 which included both 
stars.

On the one hand, in Paper I, we measured equivalent widths of
H$\delta$ and H$\gamma$ in our blue spectrum of WR\,146 to be
$W_{\lambda} \sim $0.9\AA. Comparison of these with those ($\sim
2.5$\AA) typical of mid-to-late O stars in the Walborn \& Fitzpatrick
(1990) atlas indicated that the O-star spectrum was diluted and
suggested that the continuum of the WC star was twice as bright as
that of the O companion in the blue.  On the other hand, Willis et
al. (1997) used a spectrum of lower resolution but longer wavelength
coverage than presented in Paper I to determine the WR:O light ratio
from comparison of the equivalent widths of the emission lines with
those of two single WC6 stars. They found that the O star was {\em
brighter} than the WC star, reporting a continuum light ratio (WR:O)
of 1:(2$\pm$1).

The discrepancy in the light ratios of the WC6 and O components
derived from the dilution of O star absorption lines and WC6 star
emission lines is significant and probably too great to arise purely
from the observational uncertainties.  Leaving aside the question of
whether one of the stars varied between the two observations, the
principal uncertainties in two spectroscopic light-ratio
determinations come from whether the intrinsic line strengths of the
components are indeed equal to those of the comparisons adopted. The
discrepancy could be reduced if the equivalent widths of H$\delta$ and
H$\gamma$ in the O star were weaker than the 2.5\AA\ adopted or if the
emission lines in the WC6 component were weaker than those of other
WC6 stars. Both possibilities are plausible.

The strengths of the emission lines in WR\,146 may well be
atypical for its WC6 type: both Eenens \& Williams (1992) and Willis
et al. found anomalously low C/He abundances for this star. On the
other hand, the strengths of H$\gamma$ and H$\delta$ in O-type stars
do depend on spectral type and luminosity class, which are not known
directly for the companion in WR\thinspace146. We note that the
``O8.5V'' adopted by Willis et al. comes not from the star's spectral
lines but from its luminosity inferred from that of the WC6 star and
the continuum light ratio. We therefore re-observed the blue spectrum
of WR\thinspace146 with the William Herschel Telescope in an attempt
to determine the spectral type and, hopefully, luminosity of the
early-type companion to the WR star. These data, together with the
photometry by Niemela et al., will give us a better idea of the
intrinsic properties of the companion for comparison with the radio
observations.

We also re-observed WR\thinspace146 at radio wavelengths, extending
the frequency coverage of Paper I, to search for the wind flux from
the companion, and study the characteristics of the two previously
observed radio components with the aim of furthering our
understanding of the colliding wind phenomenon.

\begin{figure}
\vspace{8.0cm}
\includegraphics{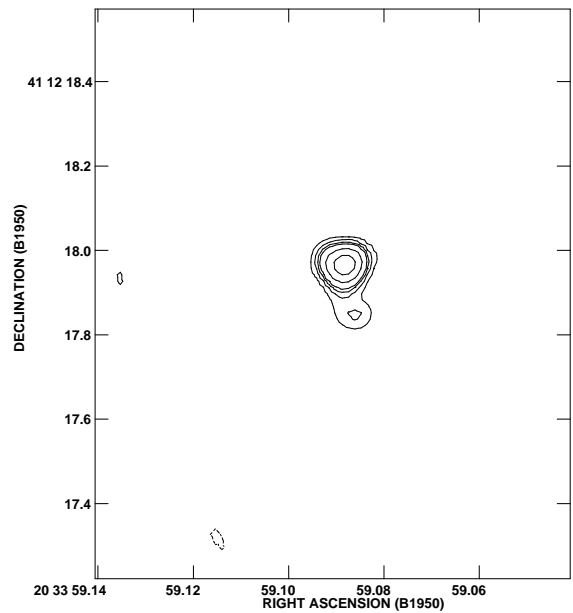}
\caption[]{ a) 5-GHz MERLIN observations obtained on 1992 December 26
The $1\sigma$ rms uncertainty is $250\mu$Jy, with contour levels at
$-3,3,6,9,12,24,48\sigma$. The synthesized beam is
$53\times53$~mas$^2$.}
\label{fig:MERLIN}
\end{figure}

\section{Radio Wavelength Study}

\subsection{The observations}

Multi-frequency observations of WR\,146 at 1.4, 5, 8.4, and 22
GHz were taken on 1996 October 26 using the A-configuration of the
NRAO Very Large Array (VLA).  Observations of the nearby radio-bright
quasar 2005+403 were interleaved with the observations of the target
source for phase-referenced calibration of the antennae gains. The
absolute flux scale was determined by observation of 3C48, assumed to
have fluxes of 15.970, 5.516, 3.226 and 1.174 Jy at 1.4, 5, 8.4, and
22 GHz respectively.

In addition, archival data from MERLIN at 5 GHz have been analysed. A
4-hour on-source observation at 5 GHz was obtained with MERLIN on 1992
December 26, along with frequent observations of the phase calibration
source 2005+403. Following the standard procedure for calibrating
MERLIN 5-GHz data, the flux scale was established by observation of
3C286 with an assumed total flux of 7.309Jy (7.020 Jy on the
MK2-Tabley baseline) and the unresolved source 0552+398, which had a
derived ``bootstrap'' flux of 5.786 Jy. The bootstrap fluxes for
2005+403 are given in Table \ref{tab:fluxes}.

\begin{figure}
\vspace{8.0cm}
\includegraphics{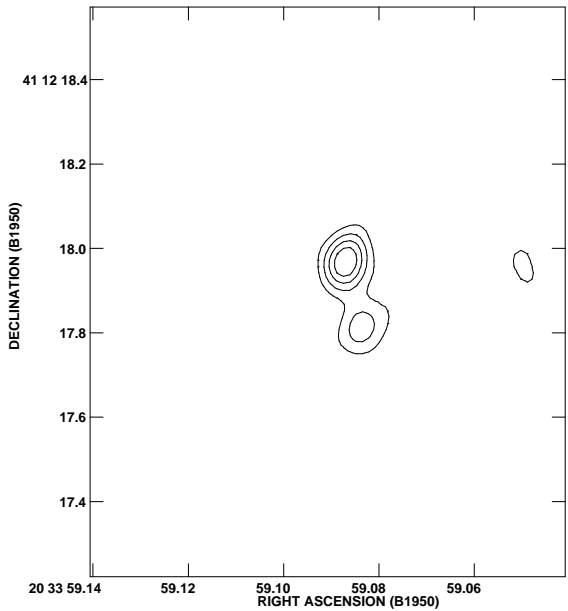}
\caption[]{VLA A-configuration observation at 22 GHz obtained on
1996 Oct 26. Two components are clearly resolved. The $1\sigma$
rms uncertainty is $0.5$mJy, with contour levels at
-3,3,6,9,$12\sigma$. The synthesized beam is $100\times75$
mas$^2$ at a position angle of $-2^\circ$. } \label{fig:22GHz}
\end{figure}

\begin{table}
\caption[]{Radiometry of WR\thinspace146}
\flushleft
\begin{tabular}{ccccc}
\hline
\multicolumn{4}{|c|}{VLA observations} \\
\hline
Epoch    &$\nu$ & N$_{22}$  & S$_{22}$ & 2005+403 \\
         & (MHz)& (mJy)& (mJy) & (Jy) \\
\hline
96/10/26 & 22460 &$10.4\pm1.0$ & $ 7.0\pm1.3$   & 2.136 \\ \cline{3-4}
         & 8435  &\multicolumn{2}{c|}{$29.8\pm0.8$} & 2.956 \\
         & 4885  &\multicolumn{2}{c|}{$37.6\pm1.0$} & 3.227 \\
         & 1465  &\multicolumn{2}{c|}{$78.4\pm0.2$} & 2.596 \\
\hline
\hline
\multicolumn{4}{c}{MERLIN observations} \\
\hline
Epoch    &$\nu$ & N$_5$& S$_5$ & 2005+403 \\
         & (MHz)     & (mJy)  & (mJy)    &  (Jy) \\
\hline
92/12/26 & 4885 &$31.4\pm0.4$ &  $2.5\pm0.3$ & 2.923\\
95/04/29 & 4885 &$28.5\pm0.3$ &  $1.6\pm0.3$ & 2.865\\
\hline
\label{tab:fluxes}
\end{tabular}

Notes: The parameters for WR\,146 at epoch 95/04/29 have been
re-determined from the original data. The flux values are slightly
higher, although consistent within the uncertainties, than the values
originally quoted in Paper I.  The quoted
uncertainties in the fluxes are the formal-fit errors. The uncertainty
in the absolute flux scale ($\sim$ a few \%) is not included and, for
at least the higher fluxes, will be an important source of uncertainty.
\end{table}

\begin{figure}
\vspace{12.6cm}
\includegraphics{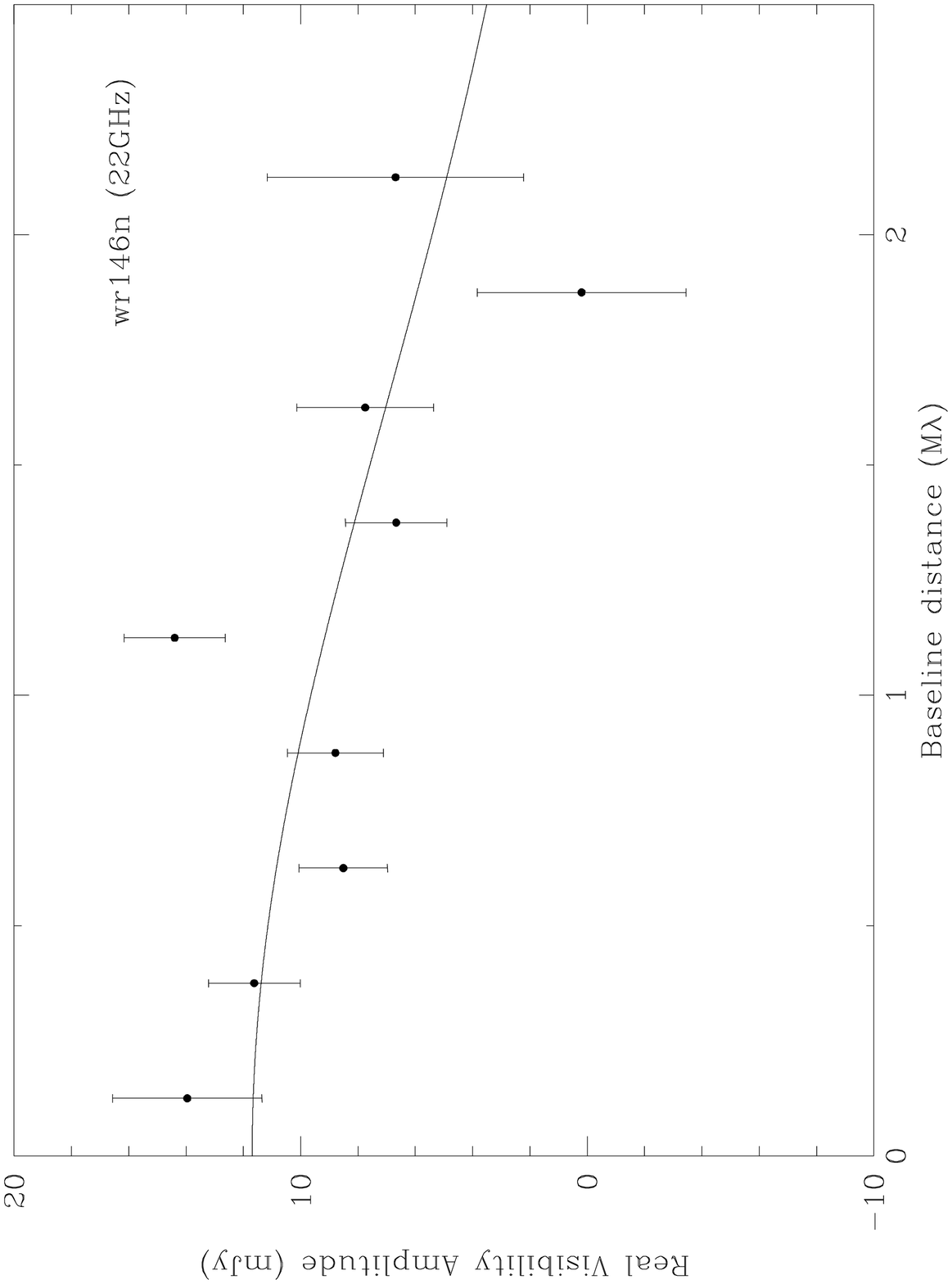}
\includegraphics{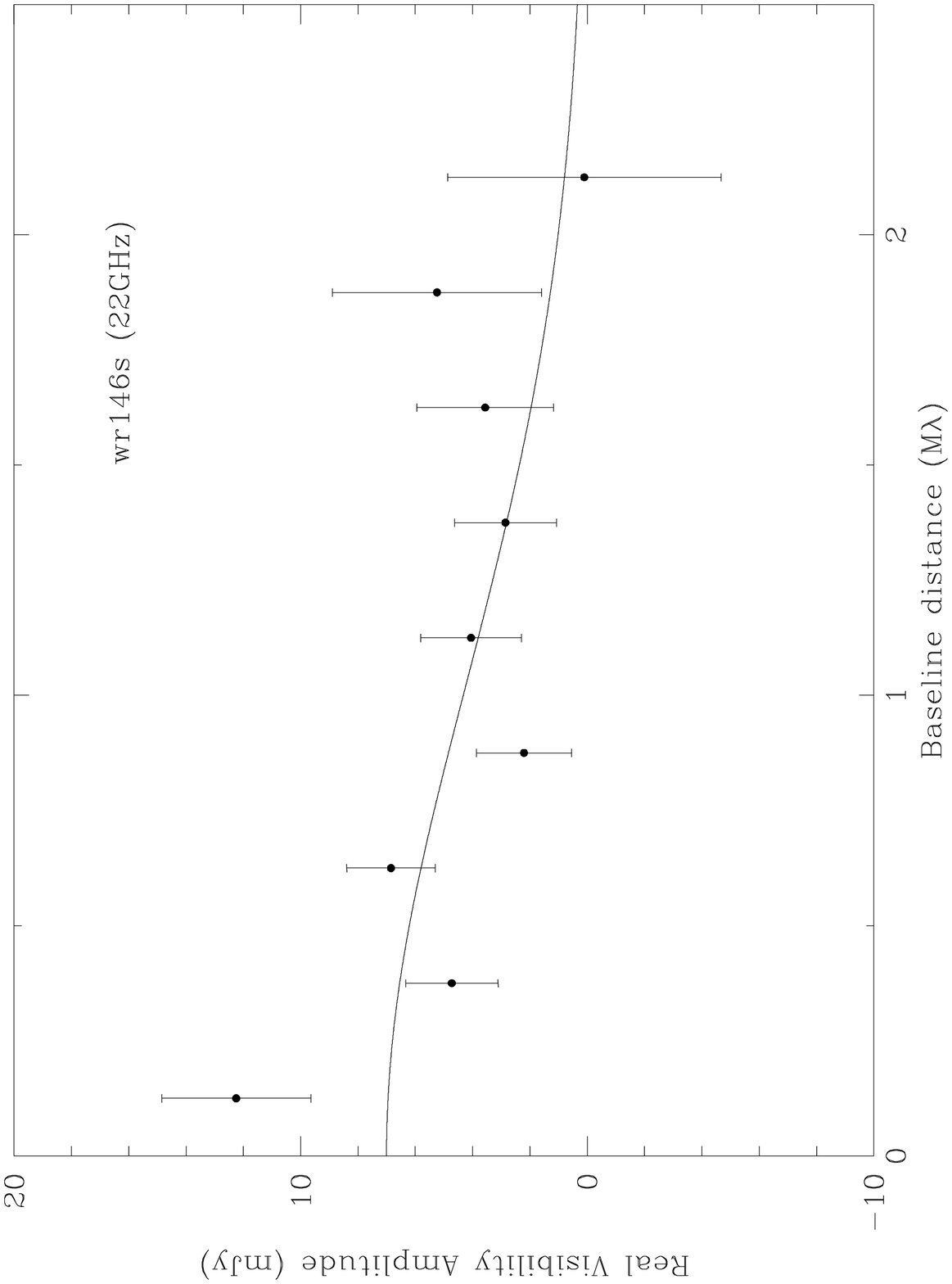}
\caption[]{The 22-GHz real visibility distributions for N$_{22}$ (top)
and S$_{22}$ (bottom). The data were binned into nine annuli in the
uv-plane, and then averaged. The averages and their uncertainties are
shown versus baseline distance. The imaginary visibility distributions
are consistent with zero. The solid lines represent weighted circular
Gaussian model fits to these data. The FWHM of the fits are $48\pm11$
and $78\pm19$~mas respectively.  It is clear from these figures that
these data do not merit more sophisticated modelling.}
\label{fig:22ghz_visibs}
\end{figure}

It is possible to check the accuracy of the derived flux scales in the
5- and 8.4-GHz observations. Comparison of the ``bootstrap'' fluxes of
2005+403 at these frequencies to those given in the University of
Michigan Radio Astronomy Observatory (UMRAO) flux database shows that, 
where closely contemporaneous observations are available, the 
agreement with UMRAO fluxes at these frequencies is within
$1\%$. Given the excellent agreement at these two frequencies, it is
reasonable to assume that the flux scales at the other observed
frequencies are good.

Aside from the initial amplitude calibration of the MERLIN
observations, the calibration and subsequent imaging of the data were
carried out using the NRAO {\sc AIPS} software package.  The complex
antennae gains were initially derived for 2005+403 and then
interpolated (phase-referenced) to the observations of
WR\thinspace146.  In addition, several iterations of phase-only
self-calibration were used to refine the antennae gains during the
MERLIN observations of WR\thinspace146 to improve the dynamic range of
the final synthesized images.

At 1.5 and 4.9 GHz the VLA data reveal an unresolved source, whereas
at 8.4 GHz the radio emission is marginally resolved. However,
WR\thinspace146 is well resolved into a double radio source at 22
GHz. The final synthesized 22-GHz image is shown in
Fig.~\ref{fig:22GHz}. We identify the northern and southern components
as N$_{22}$ and S$_{22}$. The visibility data suggest both these
components are resolved (Figure~\ref{fig:22ghz_visibs}). Gaussian
model fits to the visibilities give the diameter of the radio emitting
regions as $48\pm11$ and $78\pm19$~mas respectively (see Paper I for a
description of the method). The large relative uncertainty in these
sizes is a direct consequence of the high rms uncertainty per
visibility at 22 GHz. The fluxes from the VLA observations were
determined by Gaussian source fitting using the {\sc AIPS} routine
{\sc JMFIT}. The measured fluxes are given in Table \ref{tab:fluxes}.

The final synthesized MERLIN image at 5 GHz is shown in
Fig~\ref{fig:MERLIN}. The image reveals two components of emission,
identified as N$_5$ and S$_5$. In the MERLIN observation, the flux and
size of component N$_5$ were determined by Gaussian model fitting of
the visibility data. The derived diameter for N$_5$ is $42\pm1$~mas.
Gaussian fitting to the image data were used for the source parameters
of S given in Table 1.

Qualitatively, the 1992 MERLIN image is consistent with the 5-GHz
observation from 1995 April 29 (Paper I) though there is some evidence
that N$_5$ may have decreased in flux by $\sim3$~mJy between 1992 and
1995.  The flux of S$_5$ appears to have decreased between the two
epochs. However, the difference is only $2\sigma$, much less than the
$5\sigma$ threshold typically adopted as evidence for
variation. Therefore, we adopt the mean value of $2.0\pm0.2$ mJy for
the flux of S$_5$.  Variations in the total 5-GHz emission from
WR\thinspace146 have been observed using the WSRT over the last decade
that are attributed to variations in the non-thermal component
(Setia-Gunawan et al. 2000). Both our MERLIN 5-GHz total flux values
are consistent with the WSRT observations.  Variability could account
for the slightly higher 5-GHz flux at the epoch of the VLA
observations.

The positions of the components observed at 5 GHz and 22 GHz are
quoted in Table \ref{tab:positions}.  The absolute position of the
northern component in the various observations was deduced from the
phase-reference only calibrated images. Absolute position information
is lost during the self-calibration process, though the relative
position is preserved. These were deduced from the images shown in
Fig.~\ref{fig:MERLIN} and Fig.~\ref{fig:22GHz}.

\subsection{Identifying the radio components}
\begin{figure}
\vspace{9.4cm} 
\includegraphics{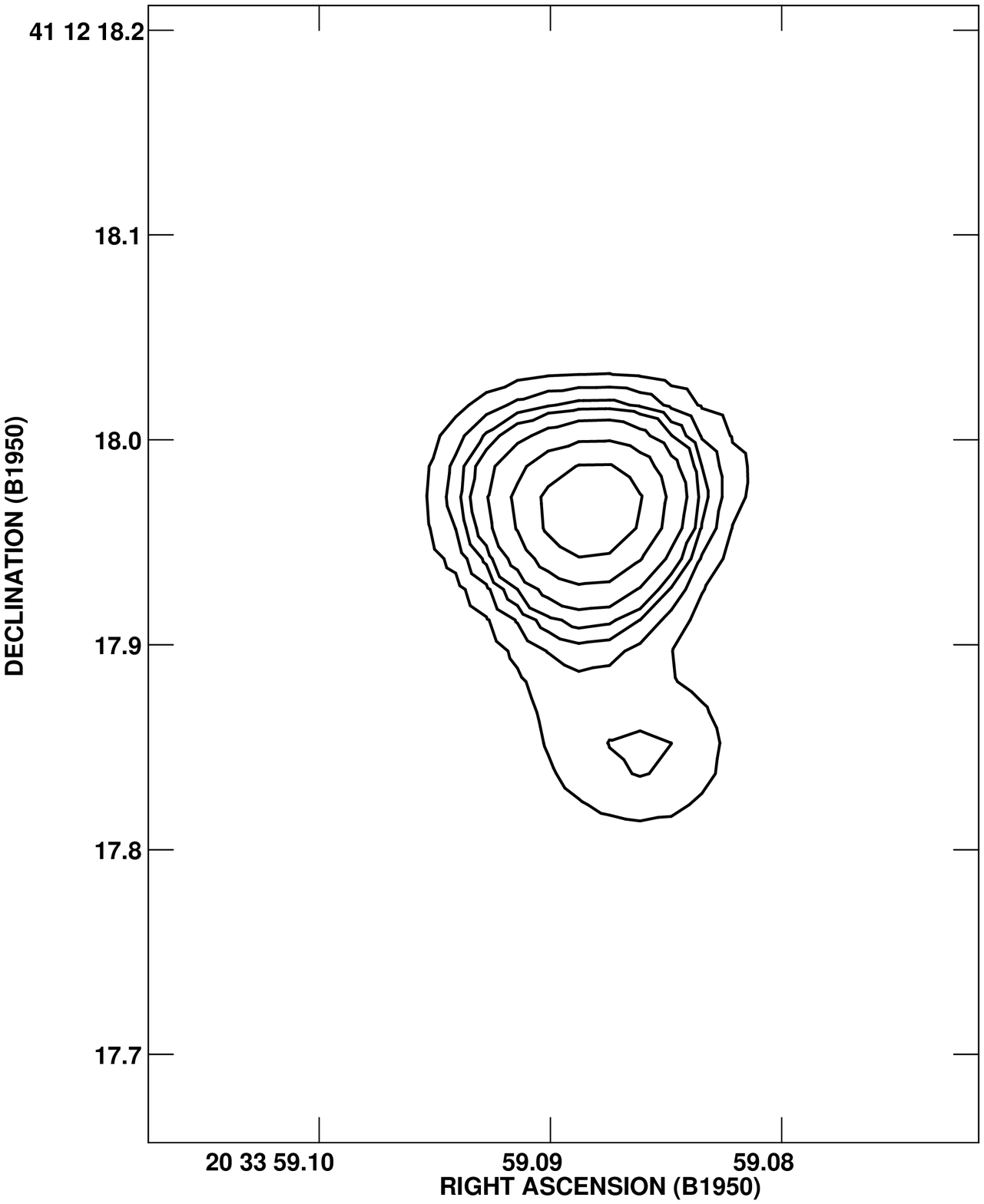}
\includegraphics{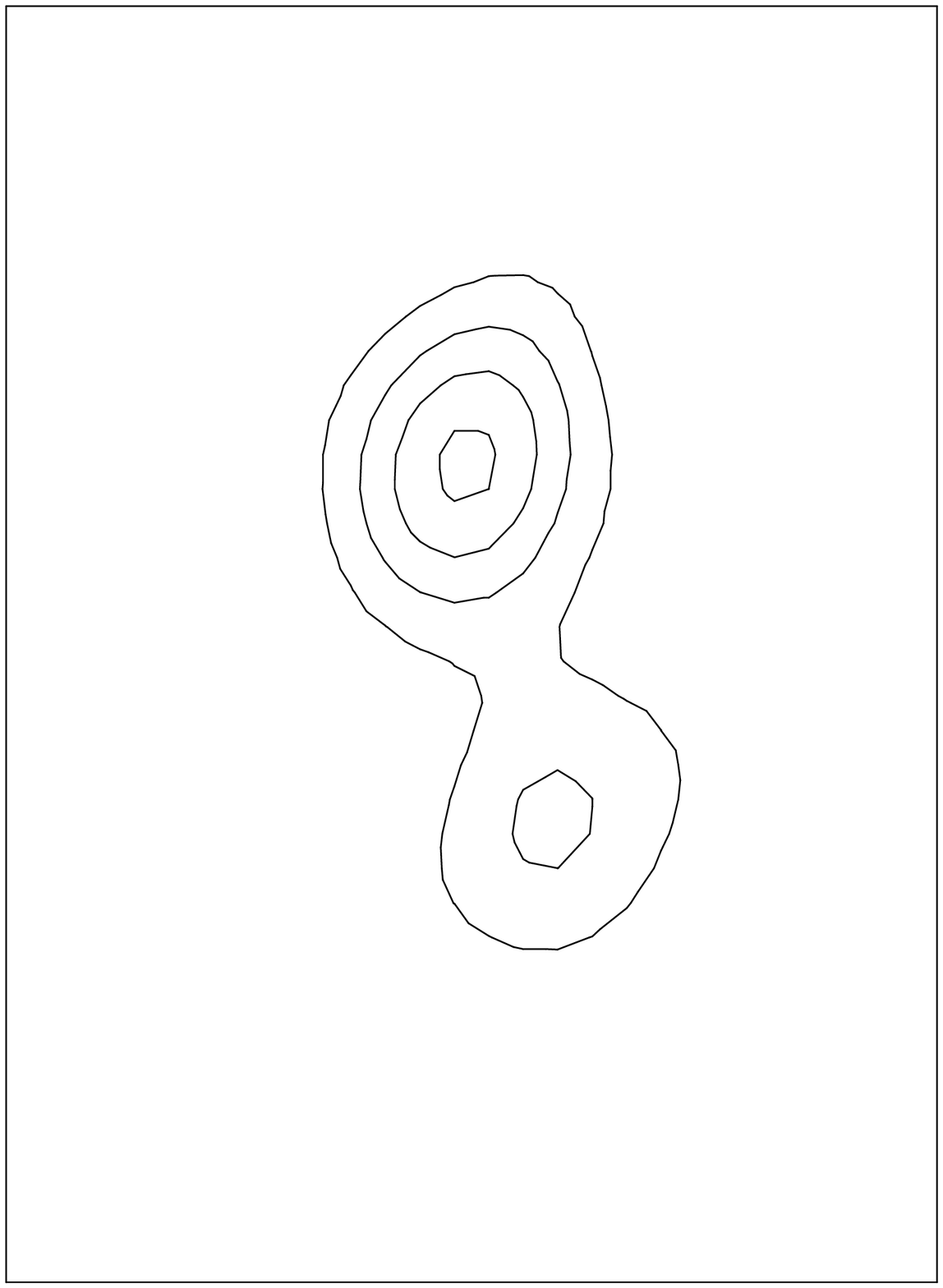}
\caption[]{Overlay of the VLA 22-GHz image (thin contours) with
the MERLIN 5-GHz image from 1992 (thick contours), assuming
S$_{22}$ and S$_5$ are coincident. Alignment was attained using
the brightest point in the two southern components. Clearly, the
northern component at 22 GHz is displaced to the north of the
non-thermal region observed at 5 GHz. We identify N$_{22}$ as the
stellar wind emission originating from the OB companion to the
WC6 star.}
\end{figure}

\begin{table}
\caption[]{J2000 Positions of emission components} \flushleft
\begin{tabular}{|l|l|l|l|l|l|}
\hline
Epoch    &Source& RA & Dec & $\Delta$ & PA \\
         &      & $20^h$ $35^m$& $41^\circ$ $22^{'}$ & (mas) & $(^\circ)$ \\
\hline
92/12/26 &N(5) & $47^s.0905$ & $44^".832$ &  & \\
         &S(5) & 47.0885 & 44.713 &  $122\pm4$ & $14\pm3$ \\

95/04/29 &N(5) & 47.0896 & 44.811 &  & \\
         &S(5) & 47.0854 & 44.705 &  $123\pm4$ & $31\pm3$ \\

96/10/26 &N(22)& 47.0997 & 44.754 &            &\\
         &S(22)& 47.0959 & 44.603 &  $162\pm8$ & $22\pm4$ \\
\hline \label{tab:positions}
\end{tabular}

Notes: the data from Paper I have been re-analysed giving the new
absolute positions quoted here. The accuracy of the absolute
positions is $\sim\pm10$~mas, due to the uncertainty in the
optical position of the phase-reference calibrator $2005+403$.
However, the relative position error is largely determined by our
ability to fit a Gaussian function to the emission from
WR\thinspace146(S).
\end{table}

The nominal resolution of A-configuration at 22-GHz observations
is very similar to that of MERLIN at 5 GHz, allowing direct
comparison of these data. At 22 GHz, the VLA observations reveal
two components N$_{22}$ and S$_{22}$, very similar in appearance
to the MERLIN 5-GHz data. However, the relative positions of the
two 22-GHz components show them to be significantly further apart
than those from the MERLIN 5-GHz data, leading to the question of
whether the 22-GHz sources are the same as those observed at 5
GHz.

The brightness temperatures of N$_{22}$, S$_{22}$ and S$_5$ are
$1.6\times10^4$~K, $\sim5000$~K and $\sim4000$~K respectively, all
consistent with an origin in a photo-ionized circumstellar
envelope where the photo-ionization equilibrium temperature is
typically $\sim10^4$~K. In contrast, this is about two orders of
magnitude lower than the brightness temperature of
$1.3\times10^6$~K for N$_5$, which clearly indicates a
non-thermal origin.

The separation of S$_{22}$ and N$_{22}$ ($162\pm8$ at a position angle
of $22\pm4^\circ$) is very close to that ($168\pm31$~mas at position
angle $21\pm4^\circ$) of the optical components (N$_{\rm O}$ and
S$_{\rm O}$) observed with the HST. There is no evidence of intrinsic
proper motion in the source at 22 GHz in recent (1999) observations of
the source (A. Fink, private communication).  The coincidence of the
optical and 22-GHz positions, the lack of evidence of intrinsic proper
motion, and the thermal nature of the emission from the two 22-GHz
components leads us to conclude that the 22-GHz components are the
stellar winds associated with the two stellar components imaged with
the HST by Niemela et al. (1988). This is the first time that the
stellar wind of the companion in a WR binary has been spatially
resolved.

In Paper I we concluded that S$_5$ was the thermal emission from
the WR-star wind. Indeed, if we take the mean flux value for S$_5$
from our two epochs of MERLIN observations, the spectral index between
22 and 5 GHz for the southern component is $+0.82\pm0.14$, consistent
with those observed in the stellar winds of other WC-type WR stars
(e.g. Williams 1996). We estimate the diameter of S$_5$ to be $\sim190$
mas from the diameter of the S$_{22}$, since angular size
$\propto\nu^{-0.6}$ for $\alpha\sim0.8$. This is approximately four
times larger than that given in Paper I.  However, we feel the diameter
presented here is a more reliable estimate.

Additionally, in Paper 1 we identified N$_5$ as non-thermal emission
from a wind-wind collision region.  If we overlay the 22-GHz and 5-GHz
data assuming two southern components originate in the WR star wind
(see Figure 4) we can see that the relative separation of the
components supports such a model, where the wind-collision region must
fall between the two stars.

\subsection{Wind parameters for both stellar components}

For a steady state, smooth, fully ionized stellar wind having a
radial $r^{-2}$ ion density distribution, the free-free radio flux
$S_\nu$ is related to the stellar-wind density i.e. $\dot M/v$ by
\begin{equation}
S_\nu= 2.32\times10^4 (\gamma g_{\nu,T_e} \nu)^{2\over 3}
\left({\dot M Z}\over{\mu v}\right)^{4\over 3} d^{-2}
\phantom{....}{\rm mJy}, \label{eqn:swf}
\end{equation}
where $\nu$ is the frequency in Hz, $d$ is the distance in kpc,
$\dot M$ is the mass-loss rate in M$_\odot$ y$^{-1}$, $v$ the
terminal velocity of the wind in km s$^{-1}$; $\gamma$, $Z$, $\mu$
and $g_{\nu,T_e}$ are respectively, the number of electrons per
ion, the mean charge per ion, the mean atomic weight and the Gaunt
free-free factor, a function of electron temperature and frequency.
It follows from (\ref{eqn:swf}) that
\begin{equation}
{\dot M \over v} \propto \left({\mu^2\over Z^2 g_{\nu,T_e}
\gamma}\right)^{1\over 2} S^{3\over 4}.
\end{equation}
Thus, the measurement of radio emission from the two stellar wind
components in the WR\,146 system leads to the density ratio of the
two winds. If the stellar winds are clumped rather than smooth, we 
can allow for this by replacing ${\dot M}$ in the above with 
$({\dot M}/\surd f)$, where $f$ is the wind filling factor, assumed 
to be constant over the radio-emitting region.

In the case of the collision of two spherical winds at terminal
velocity, the contact discontinuity between the two winds
intersects the line of centres between the stars at the point of
momentum balance. If the projected separation of the WR and OB star
is $D \cos i$, and that of the OB companion from the non-thermal
region is $r_{\rm OB}\cos i$, we can then write
\begin{equation}
r_{\rm OB}/D = \frac {\eta^{1\over 2}}{1+\eta^{1\over 2}},
\label{eqn:rdr}
\end{equation}
where $\eta$ is the ratio of WR and OB-companion wind momenta,
$(\dot M v)_{\rm OB}/(\dot M v)_{\rm WR}$. The wind-collision geometry 
and equation \ref{eqn:rdr} would also be affected by clumping, but 
this is beyond the scope of the present study. 

By defining the ratio of the OB star and WR-star wind densities, 
$(\dot M / v)_{\rm OB}/ (\dot M /v)_{\rm WR}$, as $\xi$, then
\begin{equation}
 {\dot M_{\rm OB}\over \dot M_{\rm WR}} = (\eta \xi)^{1\over2}~~~~\&~~~
 {v_{\rm OB}\over v_{\rm WR}}=({\eta\over\xi})^{1\over 2}.
\label{eqn:ratios}
\end{equation}
Knowing $\dot M_{\rm WR}$ and $v_{\rm WR}$ we can now uniquely
determine $\dot M_{\rm OB}$ and $v_{\rm OB}$ that satisfy
simultaneously both $\xi$ and $\eta$.  An attractive property of
these ratios is that they are independent of distance, typically a
very uncertain parameter for WR stars and particularly for WR\,146.

For the OB companion we will assume that hydrogen and helium are,
respectively, singly and doubly ionized, leading to values
$\gamma=1.1$, $Z=1.15$, $\mu=1.34$ (Lamers \& Leitherer 1993) and
$T_e\sim10^4$~K.  Taking the values of $\gamma=1.15, Z=1.2$,
$\mu=5.29$, $T_e=8\,000$~K from Willis et al. (1997) for the WR star,
and that the fluxes of N$_{22}$ and S$_{22}$ are those of the OB star
and the WC star respectively, implies that $\xi=0.36\pm0.19$.  From
Table 2, $D$ is $162\pm8$~mas and $r_{OB}$ is $40\pm9$~mas, so
equation~\ref{eqn:rdr} gives $\eta=0.11\pm0.03$. Thus, the ratios of
mass-loss rates and wind velocities are $0.20\pm0.06$ and
$0.56\pm0.17$ respectively.  The largest uncertainty in these ratios
arises from $\xi$. Though this requires knowledge of the relative
metallicity and ionization structure within the two winds, the values
of $Z$ and $\gamma$ are closely the same and effectively cancel. On
the other hand, the mean molecular weights of the two winds are quite
different and contributes a factor $\sim 2$ to the ratio. The relative
uncertainty in the observed fluxes provides the bulk of the
uncertainty in $\xi$.

\begin{figure*}
\vspace{10.8cm}
\includegraphics{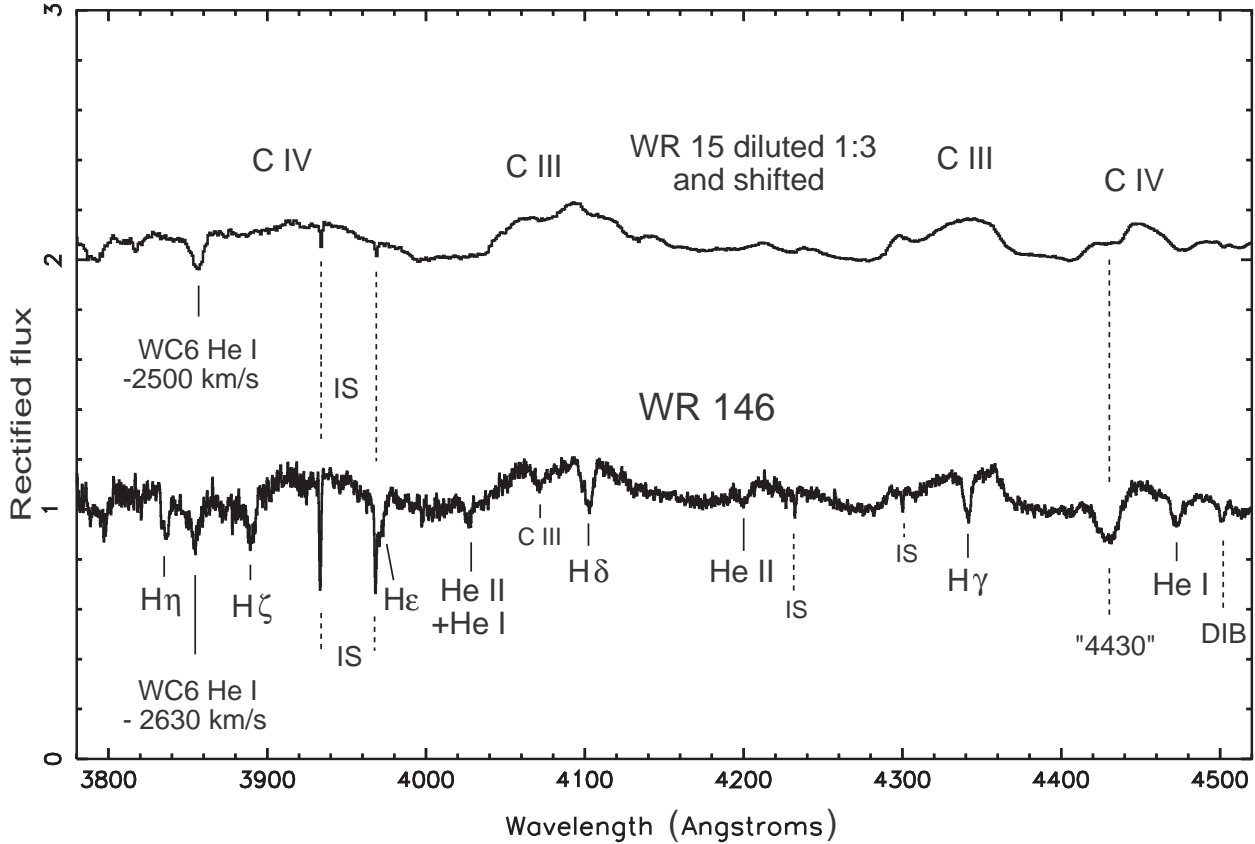}
\caption[]{Rectified blue spectrum of WR\,146 with absorption features
marked. Above it is the spectrum of another WC6 star, WR\,15. This has
been artificially diluted so as to match the emission-line spectrum
of WR\,146 as an aid to determination of the continuum in WR\,146 for
measurement of the equivalent widths of the absorption features.
Some of the diluted WR emission features are marked above the WR\,15
spectrum. \label{n146comp}}
\end{figure*}

From the deduced velocity ratio and the terminal wind velocity
measured for the WC6 star ($2900 $ km s$^{-1}$, Eenens \&
Williams 1994), we derive a terminal velocity of the OB wind of
$1600\pm480$ km s$^{-1}$.  This is consistent with those determined
for late O-type stars of any luminosity class (Prinja, Barlow \&
Howarth 1990). However, the deduced ratio (0.19) of mass-loss rates
implies a very high mass-loss rate for the OB star, irrespective of
the distance to WR\,146.  For example, if the WC6 star has a mass-loss
rate comparable to the average rate derived from radio or from
``Standard Model'' analyses of optical emission lines
($4\times10^{-5}$~M$_\odot$ y$^{-1}$, Willis 1999), that of the OB
star would be $\simeq 8\times10^{-6}$ M$_\odot $y$^{-1}$. This is
comparable to that determined for the O8f supergiant HD\,151804
(Bieging, Abbott \& Churchwell 1989, Crowther \& Bohannan 1997) and
almost $50\times$ greater than the mass-loss rates typical of
main-sequence O8 stars ($\sim 1.7 \times 10^{-7}$~M$_\odot$ y$^{-1}$,
Howarth \& Prinja 1989). This result still holds if the true mass-loss
rates of WR stars are lower than the value cited owing to clumping
since it is really the ratio of $({\dot M}/\surd f)$s that has been
determined. Only if the OB-star wind was clumped would its mass-loss
rate be reduced by a factor of $\surd f$ .

Throughout the paper we have identified the southern and northern
sources as the WR and OB star respectively. If these identifications
were switched i.e. WR star is to the north, and the OB star to the
south, then the deduced wind velocity of the OB star is greater than
$18\,000$~km s$^{-1}$ and the mass-loss rate is 1.4 times that of the
WR star. Clearly these parameters are unacceptable given our knowledge
of OB-star winds and we dismiss such a source identification as
untenable.

The mass-loss rates determined from the measured free-free wind fluxes
depend heavily on the adopted distance to WR\thinspace146. Therefore,
we first re-examine the luminosity of the system using new optical
observations.

\section{Optical Spectroscopy}

\subsection{The observations}

The spectrum of WR\,146 was taken with the blue arm of the ISIS
spectrograph on the William Herschel Telescope (WHT) on
1999 June 19.  As in Paper I, we observed in the blue, where the
relative contribution of the OB companion would be greatest, and
avoided the strong $\lambda$4650\AA\ emission feature from the
WR star. The R1200B grating and 1.1 arcsec slit gave a resolution
of 0.9 \AA.  Four integrations, each of 750s, were taken of the
star, interleaved with observations of an argon discharge tube
for wavelength calibration. The detector used was a
$4200\times2048$ EEV42 device.

We present the new spectrum of WR\,146 in Fig. \ref{n146comp} and
list absorption lines in Table \ref{opt}.  The spectrum has been
smoothed giving a resolution $\sim$ 1.0 \AA\ (FWHM), as measured
from the widths of the narrow interstellar CH, CH$^{+}$ and
Ca\,{\sc i} lines. The FWHM of the $\lambda$3933\AA\ Ca\,{\sc ii}
line is 1.6 \AA, the greater width coming
from optical depth effects (cf. Jenniskens \& D\'esert 1994).
The well developed interstellar spectrum is consistent with the
heavy reddening suffered by WR\,146.

The broad undulations in the continuum are mostly emission lines
from the WC6 star, diluted by the light from the companion. The
emission lines in WR\,146 are unusually broad for its WC6
subtype (cf. Eenens \& Williams 1994, Willis et al. 1997), making
it easier to measure the absorption spectrum of the companion.
The absorption line widths are typically $\sim$ 5\AA\ FWHM,
comparable for example to those of the O9.5II star $\delta $ Ori
(cf. Voels et al. 1989), which has a relatively high $v \sin i$
(144 km\,s$^{-1}$, Howarth et al. 1997).

There is one absorption feature from the WC6 star; the P-Cygni
absorption component of the $\lambda$3889\AA\ He\,{\sc i} line,
blue-shifted by 2630 km\,s$^{-1}$ to 3854.5\AA.  This line arises from
the same metastable level as the $\lambda $10830\AA\ line from which
Eenens \& Williams (1994) derived a terminal wind velocity
(v$_{\infty}$) of 2900 km\,s$^{-1}$ by profile fitting and, like the
$\lambda $10830\AA\ line, is expected to form in the outer regions of
the WC6 wind.  These results are consistent with the
v$_{\infty}=2700$~km\,s$^{-1}$ measured by Willis et al.  from the
[Ne\,{\sc iii}] $\lambda $15.5-$\mu $m fine-structure line, also
formed in the outer wind.

\begin{table}
\centering

\caption{Measured wavelengths (\AA) and identifications of
absorption features in the spectrum of WR\,146.  The ``$\lambda$
lab'' values for the interstellar features come from Jenniskens
\& D\'esert (1994). \label{opt}}
\begin{tabular}{lll}
\hline $\mathbf{\lambda }$ obs & $\mathbf{\lambda }$ lab &
Identification (and equivalent width, W$_{\lambda }$) \\ \hline
3835.4 & 3835.4 & H$\eta $ \\
3854.5 & 3888.6 & He\,{\sc i} from WC6 star at --2630 km s$^{-1}$ \\
3889.6 & 3889.0 & H$\zeta $ (?+ 3888.6 He\,{\sc i} 2s-3p) \\
3933.4 & 3933.4 & Interstellar Ca\,{\sc ii} \\
3968.3 & 3968.5 & Interstellar Ca\,{\sc ii} \\
3970.4 & 3970.1 & H$\varepsilon $ \\
4025.9 & 4026.2 & He\,{\sc i} 2p-5d  + 4025.6 He\,{\sc ii} \\
4071.1 & 4070.3 & C\,{\sc iii} 4f-5g \\
4089.6 & 4088.9 & Si\,{\sc iv} W$_{\lambda }$ = 0.06$\pm $0.03\AA \\
4102.6 & 4101.8 & H$\delta $ (+4100.0 He\,{\sc ii}) \\
4200.8 & 4199.8 & He\,{\sc ii}; W$_{\lambda }$ = 0.30$\pm$0.03\AA \\
4226.6 & 4226.7 & Interstellar Ca\,{\sc i} \\
4232.3 & 4232.5 & Interstellar CH$^{+}$ \\
4299.9 & 4300.3 & Interstellar CH \\
4340.9 & 4340.5 & H$\gamma $ (?+4338.7 He\,{\sc ii});
W$_{\lambda }$ = 0.70$\pm $0.05\AA  \\
4430 & 4428.9 & DIB ``$\lambda $4430'' \\
4472.1 & 4471.5 & He\,{\sc i} 2p-4d; W$_{\lambda }$ = 0.50$\pm $0.04\AA \\
4501.4 & 4501.8 & DIB \\ \hline
\end{tabular}
\end{table}

The other absorption lines (Table \ref{opt}) are formed in the
OB-companion star. Each of the hydrogen lines is blended with the
nearby member of the Pickering series of He\,{\sc ii} but
the relative weakness of the unblended odd-numbered Pickering
line at 4200\AA\ and the closer proximity of the hydrogen-line
wavelengths to the observed wavelengths suggests that the
He\,{\sc ii} contribution is relatively small.
Helium\,{\sc i} is represented by the $\lambda\lambda$ 4472 and
4026\AA\ lines; the latter is blended with He\,{\sc ii} and is
therefore not very useful for diagnostic purposes.
Other He\,{\sc i} lines in the Walborn \& Fitzpatrick (1990) atlas
of hot-star spectra are not seen in our spectrum. The relative
strength of the $\lambda$ 4070\AA\ C\,{\sc iii} triplet is a
little surprising but the continuum here is uncertain as it falls
between two emission features from the WC6 spectrum.

In order to establish the continuum in the regions of each of the
absorption lines so that we could measure their equivalent
widths, we compared our rectified spectrum with that of another
WC6 star, WR\,15 (HD 79573), artificially diluted so as to match
the spectrum of WR\,146. The spectrum of WR\,15 was observed
with the 1.9m telescope at the South African Astronomical
Observatory and will be discussed elsewhere.
As a by-product of this matching, we estimated the dilution of
the WC6 spectrum in WR\thinspace 146, finding OB:WR $\approx $ 3,
consistent with the ratio, OB:WR $=$2$\pm $1, found by Willis et
al. (1997) from stronger emission lines between 4660\AA\ and
6560\AA. We are chary of putting too much weight on this result
as it depends heavily on the rectification process and the
presumption that the WC6 emission lines in WR\,146 have
the same strengths as those in WR\,15. Instead, we
concentrate on the absorption lines, to be discussed below.

\subsection{Spectral type and luminosity of the OB companion}

To investigate the nature of the OB companion, we begin by using the
helium-line ratios to estimate its spectral type.  Our optical spectrum
includes contributions from both N$_{\rm O}$ and S$_{\rm O}$ but the
ratios of the absorption lines in the companion will be equal to those
measured from our combined-light spectrum, and independent of the
dilution by the WR star, provided that the dilution does not change
significantly over the wavelength range of the spectrum.

Formal O-star spectra classification is based on the ratio of the
$\lambda$4541\AA\ He\thinspace{\sc ii} and $\lambda $4472\AA\
He\thinspace{\sc i} lines. Since the former was not available, we used
the $\lambda $4200\AA\ He\thinspace{\sc ii} line from the same series.
We formed an empirical calibration of
$\lambda$4472\AA/$\lambda$4200\AA\ ratios against spectral type using
the line strengths measured for a large ($\sim $ 100) sample of O
stars by Conti \& Alschuler (1971) and Conti (1973).  Using this
calibration, the ratio of our measured equivalent widths (Table
\ref{opt}) indicates a type of O8, with an uncertainty of half a
subclass. The type is close to that inferred by Willis et al. from
their OB:WR light ratio and an average luminosity for the WC6 star;
but this agreement is fortuitous.

The luminosity class criterion adopted by Conti \& Alschuler for the
middle and late-type O stars was the ratio of the Si\,{\sc iv}
$\lambda $4089\AA\ and He\,{\sc i} $\lambda $4143\AA\ lines.
Unfortunately, the silicon line is too weak (W$_{\lambda } \sim $
0.06\AA) to measure with confidence and the $\lambda $4143\AA\ line is
not seen at all.  Another significant He\,{\sc i} line that is
apparently missing is the $\lambda $4388\AA\ singlet line. The
strength of this relative to that of the $\lambda $4472\AA\ He\,{\sc
i} line (``singlet to triplet ratio'') has long been known to be
sensitive to luminosity owing to the relative overpopulation of the
He\,{\sc i} 2$^{3}P^{o}$ state in extended atmospheres (e.g. Voels et
al. 1989 and references therein).  Using the equivalent-width
measurements by Conti \& Alschuler (1971) and Conti (1973), we
examined the ratios, W$^{\prime }$= W$_{\lambda }$($\lambda
$4388)/W$_{\lambda }$($\lambda $4472), formed from their observations
of the O7.5, O8 and O8.5 stars in their sample.  This ratio was found
to be a strong diagnostic for Of stars.  All the O7.5--8.5 stars
having W$^{\prime } < 0.25$ are Of stars, most of them
supergiants. Measurement of the $\lambda $4472\AA\ line and
examination of our spectrum in the region of the $\lambda $4388\AA\
line indicates that W$^{\prime }$ $\ll $ 0.2, strongly suggesting that
the O8 star in WR\,146 is an Of star, and possibly a supergiant.
Unfortunately, the classical Of diagnostic emission features C\,{\sc
iii} $\lambda\lambda $4630--41\AA\ and He\,{\sc ii} $\lambda $4686\AA\
fall within the strong $\lambda $ 4650\AA\ C\,{\sc iii-iv}/He\,{\sc
ii} emission feature from the WC6 star, which extends from 4620\AA\ to
4690 \AA\ in WR\thinspace146 (Dessart et al. 2000). This increases the
dilution of the OB spectrum to $\sim$ 1:4 in this critical region and
we cannot make a formal Of classification.

Further circumstantial support for a high luminosity for the OB star
comes from comparison of the $\lambda $4472\AA\ He\,{\sc i} to
H$\gamma $ ratio with those derived from the Conti \& Alschuler and
Conti (1973) datasets. On the other hand, we do not observe emission
lines corresponding to the `unidentified' features observed in
O6--O9.7 supergiants, including the O8If star HD\,151804 (Crowther \&
Bohannan 1997), at 4486\AA and 4504\AA. Also, the weakness of the
Si\thinspace{\sc iv} $\lambda$4089\AA\ line is surprising given the
other evidence for high luminosity, but not unique: the Si\,{\sc iv} 
line is barely visible in the spectrum of the O7Iaf star Sanduleak 80 
(Walborn \& Fitzpatrick 1990). Although we cannot assign a luminosity 
class, the spectrum points to an extended atmosphere and high luminosity, 
but not as high as that of HD\,151804.

\subsection{Spectral line dilution and the HST photometry}

With a better idea of the nature of the O8 star, can we reconcile the
measured absorption-line strengths in our composite spectrum with the
observations by Niemela et al. (1998)? We follow Paper I and Niemela
et al. by assigning the northern optical image (N$_{\rm O}$) to the
O8 star on the basis of its closer proximity to the non-thermal radio
source and the assumption that its stellar wind has a lower momentum
than that of the WC6 star. If we assume that the spectrum of S$_{\rm O}$
is that of a WC6 star, we can estimate the intrinsic equivalent
widths in the O8 spectrum from those observed in our combined-light
spectrum and Niemela et al.'s photometry of S$_{\rm O}$ and N$_{\rm O}$.
We find 1.4\AA\ for the intrinsic equivalent width of H$\gamma $, which 
is closer to the average of those measured by Conti (1973) for O8f stars 
(1.3$\pm $0.3\AA) than for `non-Of' O8III--V stars (2.5$\pm $0.2\AA), 
another indication of high luminosity properties for the OB star.
Our earlier view (Paper I) that the WC6 star was brighter than the OB
star was based on the erroneous assumption of intrinsic absorption-line
strengths appropriate for a main-sequence OB star.

The observation of approximately equal B magnitudes for N$_{\rm O}$ and
S$_{\rm O}$ is not consistent with the light ratios derived from dilution
of the emission-line spectrum of the WC6 star by Willis et al. (1997),
and supported by our estimate in Section 2.1. This could be resolved also
if the emission lines in the WC6 component were 30--40\% weaker than
those typical of WC6 stars. This is plausible; as already noted, the
star appears to have an abnormally low C/He abundance ratio and the
dilution analysis by Willis et al. depends heavily on C\,{\sc iii} and
C\,{\sc iv} lines.  Therefore, the very different light ratios originally
derived from combined-light spectra in Paper I and by Willis et al. can be
reconciled with a model in which WR\,146 comprises two stars of 
approximately equal brightness in the blue and relatively weak lines, 
those of N$_{\rm O}$ being consistent with other indicators of its high 
luminosity.

\subsection{The nature of the northern component}

The H$\gamma $ strength (together with HST photometry), the He\,{\sc i}
singlet-to-triplet line ratio, and the very high mass-loss rate all
point to an extended atmosphere for N$_{\rm O}$. If it is single, it
cannot be a main-sequence star but must be luminous. Alternatively, it
may itself be an unresolved binary comprising an O8 star with a third
component.

If N$_{\rm O}$ was a single, luminous star, we can estimate its absolute 
magnitude from those of O8I/f and O8III stars ($M_{V}=-6.5$ and $-5.5$, 
Conti 1988; Vacca, Garmany \& Shull 1996). For example, combination of 
an intermediate luminosity, $M_{V}=-6.0$, which might be appropriate 
given the apparent absence of the `unidentifed' 4486\AA\ and 4504\AA\ 
emission lines, with the photometry by Niemela et al. would give 
$M_{V}=-6.2$ for S$_{\rm O}$ and a combined $M_{V}=-6.9$ for the 
WR\,146 system. The implied luminosity of the WC6 star, $M_V = -6.2$  
or $M_v\simeq -6.0$ on the narrow-band $ubv$ system used for WR stars, 
is significantly greater than those of other WC6 stars (e.g. mean 
$M_v=-4.1\pm 0.4$, van der Hucht, in preparation), but this may be 
another manifestation of the anomalous nature of the WC6 star in WR\,146.  

Adopting a reddening of $E_{B-V}=2.7$, consistent with Paper I and 
Willis et al. (1997), the combined $M_V$ implies a distance modulus 
$m-M=11.2$, greater than our distance in Paper I ($m-M=10.4$) but 
consistent with membership of the Cyg OB2 Association ($m-M=11.2$, 
Massey \& Thompson 1991; Torres-Dodgen, Tapia \& Carroll 1991). 
Given the angular distance ($\simeq 0.5^\circ$) of WR\,146 from the 
centre of the association, stronger evidence is needed to confirm 
membership. 

At this distance (1.7 kpc), the mass-loss rates derived for the WC6 and 
O8 stars are very high: $1.3 \times 10^{-4}$ M$_\odot $y$^{-1}$ and 
$2.5 \times 10^{-5}$ M$_\odot $y$^{-1}$ respectively. The latter is three 
times that of HD\,151804, whereas the luminosity of N$_{\rm O}$ appears 
to be lower. Such a system would have stronger emission lines that those 
in HD\,151804, contrary to observations (e.g. 4486\AA\ and 4504\AA\ or 
H$\alpha $ in Dessart et al. 2000, fig 4 compared with Crowther \& 
Bohannan 1997), so the single high-luminosity model for N$_{\rm O}$ is 
problematic. 

If we consider N$_{\rm O}$ to be an unresolved binary in which some of
the mass-loss is contributed by a third component, the latter is
unlikely to be another O-type star. Since $\dot M \propto L^{1.7}$ for
O stars (Howarth \& Prinja 1987), we expect the mass-loss rate of an
O+O system to be lower than that of a single star having the same
total luminosity.  Also, the intrinsic strengths of the H$\gamma $
lines from two lower luminosity stars are likely to be greater, in
conflict with our combined-light spectrum and the HST photometry.

A model wherein the third component is another WC star suffers neither 
of these problems. Its spectral subtype would have to be similar to 
that of S$_{\rm O}$, or it would have to be fainter, not to have been 
detected in the combined-light spectra. Its presence could account for 
some of the anomalous properties (e.g. low C/He) of the WC6 star reported 
in the studies referred to. 
It could also modulate the wind of the O8 star flowing towards the wind 
interaction region, providing a mechanism for the 3.38-y variability in 
the non-thermal emission reported by Setia Gunawan et al. (2000).
To examine this model, separate spectra of N$_{\rm O}$ and S$_{\rm O}$,  
e.g. with an adaptive optics system, are needed to see whether N$_{\rm O}$ 
has a WC companion (or Of emission lines) and to determine its spectrum 
without dilution.

\section{Conclusions}

From high-resolution observations with the VLA at 22 GHz and with
MERLIN at 5 GHz we have observed all three components of the WR\,146
system: the OB and WC6 stellar winds and the non-thermal source where
they collide.  The source geometry and ratio of stellar wind fluxes
allow us to determine the ratios of mass-loss rates and wind
velocities independent of distance to the system. From these ratios
and the observations of the WC6 star, we derive the wind velocity of
the OB star to be $1600\pm480$ km\,s$^{-1}$ and its mass-loss rate to
be one quarter that of the WC6 star.  If the WC6 star has an
``average'' WR-star mass-loss rate of $\sim
4\times10^{-5}$~M$_\odot$\,y$^{-1}$, that of the OB star would be
$\sim 8\times 10^{-6}$~M$_\odot$\,y$^{-1}$, suggesting a very high
luminosity object.

Support for high luminosity comes from the optical spectrum of
WR\,146, which includes both stars. This shows absorption-line ratios
formed in the OB companion suggesting it to be a high-luminosity O8
star. If it is a single star, the inferred luminosity places WR\,146
at the distance of the Cyg OB2 association. This gives an anomalously
high luminosity for the WC6 star but, given its other anomalies, does
not rule out this distance.  However, the mass-loss rates determined
using this distance ($1.3 \times 10^{-4}$ M$_\odot $y$^{-1}$ for the
WC6 star and $2.6 \times 10^{-5}$ M$_\odot $y$^{-1}$ for the O8 star)
are awkwardly high and the latter is probably inconsistent with the
spectroscopy.  Many, if not all, of the observations could be
explained if the companion was itself a binary comprising an O8 star
and another WC star. This needs to be tested by separate spectra of
the two visual components of WR\thinspace146. The presence of an
unresolved companion to the O8 star could modulate its wind so as to
cause the 3.38-y variability in the non-thermal emission reported by
Setia Gunawan et al. (2000).

We may not have reached a firm conclusion as to the nature of the
stellar companion(s) to the WC star from the new radio and optical
observations. However, the ability to study all three radio components
of the WR\thinspace146 system separately will ensure that this system
becomes an archetype for studying the wind-collision phenomenon.

\section*{Acknowledgements}
This research has made use of data from MERLIN, a UK national facility
operated by the University of Manchester, and the William Herschel
Telescope, operated on Observatorio del Roque de los Muchachos, La
Palma, by the Isaac Newton Group, both on behalf of Particle Physics
and Astronomy Research Council; the National Radio Astronomy
Observatory Very Large Array, a facility of the National Science
Foundation operated under cooperative agreement with Associated
Universities, Inc; the University of Michigan Radio Astronomy
Observatory supported by the National Science Foundation and the
University of Michigan; the South African Astronomical Observatory;
and the SIMBAD database, operated at the CDS, Strasbourg, France. The
authors are indebted to the referee, Ian Howarth, for his detailed
critique of the original manuscript and suggestions for improvements
that sharpened our thinking; and to Amy Fink for sharing her
observations prior to publication. SMD is indebted to the National
Research Council of Canada, Dominion Radio Astrophysical Observatory
for supporting this research.

\end{document}